\documentclass{article}
\usepackage{ijcai22}

\usepackage{times}

\usepackage{soul}
\usepackage{url}
\usepackage{color}
\usepackage[hidelinks]{hyperref}
\usepackage[utf8]{inputenc}
\usepackage[small]{caption}
\usepackage{graphicx}
\usepackage{amsmath}
\usepackage{booktabs}
\title{Text/Speech-Driven Full-Body Animation}

\author{
Wenlin Zhuang\footnote{Equal Contributors.}\and Jinwei Qi\footnotemark[1]\and Peng Zhang\and Bang Zhang\And Ping Tan\\
\affiliations
XR Lab, Alibaba Group\\
\emails
\{zhiyu.zwl, jinwei.qjw, futian.zp, zhangbang.zb, xingye.tp\}@alibaba-inc.com
}

\begin{document}

\maketitle

\begin{abstract}
Due to the increasing demand in films and games, synthesizing 3D avatar animation has attracted much attention recently. In this work, we present a production-ready text/speech-driven full-body animation synthesis system. Given the text and corresponding speech, our system synthesizes face and body animations simultaneously, which are then skinned and rendered to obtain a video stream output. We adopt a learning-based approach for synthesizing facial animation and a graph-based approach to animate the body, which generates  high-quality avatar animation efficiently and robustly. Our results demonstrate the generated avatar animations are realistic, diverse and highly text/speech-correlated.
\end{abstract}

\section{Introduction}
Automatic synthesis of  3D avatar animation is useful in a variety of applications including entertainment applications (such as film and game), educational applications, and so on. Conventionally, full-body animation is always accompanied by text and speech. In reality, the face and body begin to move when the  actor begins to speak, perform actions according to the speech, and actors 
can perform some semantic body movements to express specific text semantics (e.g., number, orientation). The complicated mapping relationship between full-body movement and text/speech has prompted researchers to investigate text/speech-to-animation synthesis.

%% 
%Text/speech-driven avatar animation aims to generate both face and body movements to match the given text and speech. The main challenge is the mapping between text/speech and animation is one-to-many, due to the fact that people can preform various movements when talking. Previous research mainly concentrate on a single part, either the face animation~\cite{karras2017audio,wiles2018x2face,oh2019speech2face,chen2019hierarchical} or the body animation~\cite{alexanderson2020style,shlizerman2018audio,ginosar2019learning}, which reduces the expressiveness of the avatar. 
%In the previous work of face animation, the lip movement as well as upper face expression are mostly predicted from the given speech fragments by an end-to-end learning manner, but taking speech information as the only input limits the performance of face animation. On the one hand, the speech of real scene may contains noise, which affect the recognition of speech content leading to incorrect lip movement. On the other hand, the synthetic speech is lack of emotional change that leads to poor expression performance. Furthermore, the textual script can also provide rich content information for more expressive animation that is missed by the existing works. 
Text/speech-driven avatar animation aims to generate both face and body movements to match the given text and speech. %The main challenge is the mapping between text/speech and animation is one-to-many, due to the fact that people can preform various movements when talking. （futian: Deleted ）
Previous research mainly concentrate on a single part, either the face animation~\cite{karras2017audio,wiles2018x2face,oh2019speech2face,chen2019hierarchical} or the body animation~\cite{alexanderson2020style,shlizerman2018audio,ginosar2019learning}, which reduces the expressiveness of the avatar. 
%To tackle this problem, our algorithm animates both face and body with the same inputs. 
Recently, some researchers~\cite{hu2021text,hu2021virtual} jointly synthesized face and body animation, and they adopted logic triggered by fixed motion segments, which makes it difficult to synthesize diverse and co-speech animations.
In the previous work of face animation, the lip movement as well as upper face expression are mostly predicted from the given speech fragments by an end-to-end learning manner. The speech data is either collected real voice or generated by text-to-speech algorithm(TTS).  Speech of real scene may contain noise, which affects the recognition of speech content leading to incorrect lip movement. While synthetic speech lacks of emotional change, which leads to poor expression performance. Furthermore, taking speech as the only and missing the text information may limit the algorithm performance. While text provides rich content information for more expressive animation.
For the body animation, 
most researchers used deep learning-based approaches to achieve end-to-end synthetic body movements. However, these methods  tend to generate the average pose of all possible target poses for each frame due to the inherent one-to-many mapping between speech and motion, resulting in boring motions. More importantly, their works only focus on phonetic features while ignoring semantic information, which is a crucial information for motion-speech synergy.

In this paper, we propose an end-to-end framework to simultaneously synthesize face and body animation. Face animation adopts a learning-based method to generate lip movement and facial expression, which constructs a multi-pathway transformer network to jointly model the speech information as well as phoneme labels from textual scripts. Besides, we also adopt text analysis to extract semantic key words to perform expression fusion. In the branch of body animation synthesis, we adopt a motion graph-based retrieval method~\cite{kovar2008motion,ferstl2021expressgesture,chen2021choreomaster}. This effectively avoids average pose when synthesizing. In addition, we propose two synthesis rules to make the whole synthesis process fully controllable. First, special semantic text requires corresponding semantic motions. Second, the rhythm of non-semantic motion needs to be aligned with the phonetic rhythm. This not only ensures that the synthesized motions are highly speech-correlated, but also the semantic movement greatly enhances the expressiveness of the animation. Furthermore, our method can synthesize diverse movements under the rules, and the demonstration video is shown in \url{https://youtu.be/MipiwU3Em_8}.

\begin{figure*}
\centering
  \includegraphics[width=0.99\textwidth]{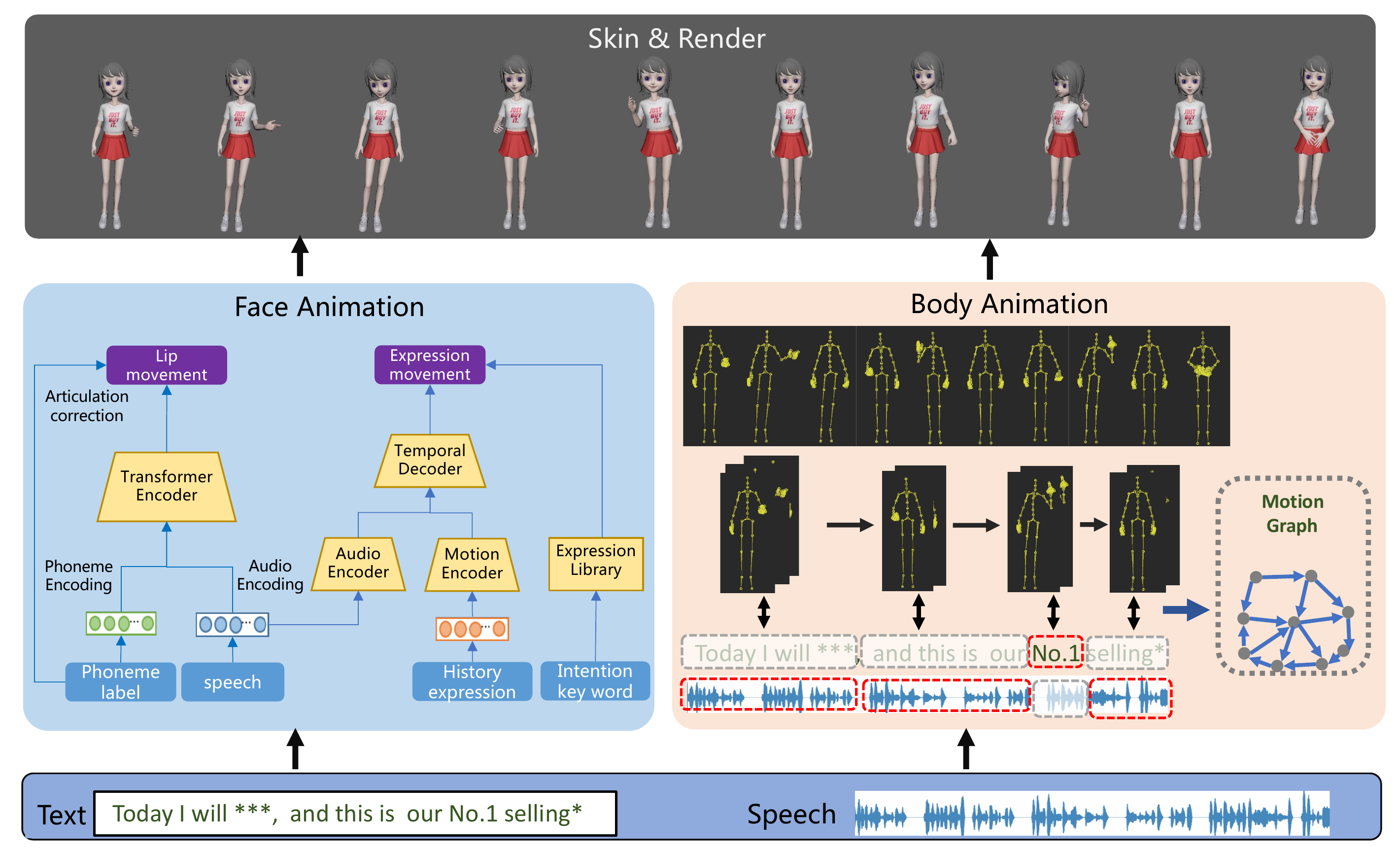}
  \caption{Overview of our proposed framework.}
  \label{fig:framework}
\end{figure*}

\section{Framework}
Here we describe our framework towards the problem of text/speech conditioned face and body animation as shown in Figure \ref{fig:framework}. Given a section of text and speech, the human face and the body are synthesized through two branches respectively. One branch adopts a learning-based method to synthesize lip motions and expressions, while the other branch uses a database retrieval-based method to synthesize skeleton motion. Full-body animation is then obtained through skinning and rendering.

\subsection{Face Animation}
We first introduce our proposed learning-based face animation approach, where the facial movements are divided into two major parts including lip movement in the lower face and diverse expressions in the upper face. We construct a multi-pathway framework to generate movements of two facial parts respectively. Specifically, we use the facial motion capture device to collect 3 hours human talking data with diverse expressions. We record both video data as well as 3D face parameter sequences under the definition in ARKit with 52 blendshapes. %, which are further used to drive the facial movements of 3D avatar. 
%\footnote{The demonstration video: \url{https://youtu.be/MipiwU3Em_8}}
\textbf{Lip movement generation.}
To generate accurate and synchronous lip movements, we construct a cross-modal transformer encoder to utilize both speech and textual information. For the modeling of textual information, we extract phoneme alignment annotation according to the speech and textual scripts by time alignment analyzer such as Montreal Forced Aligner toolkit, denoted as $Ph=\{ph_t\}^T_{t=1}$ where $ph_t$ is the phoneme label at the $t$-th frame. 
While for the speech signals, we concatenates Mel Frequency Cepstral Coefficients (MFCCs) features and Mel Filter Bank (MFB) features denoted as $Au=\{au_t\}^T_{t=1}$. 

%Our proposed cross-model 
The proposed transformer encoder takes a sequence of concatenated phoneme embedding and audio features as input with a window size of 25 frames, whose duration is 1 second. The transformer encoder can effectively model the temporal context information with a multi-head self-attention mechanism across different modalities. We minimize the following loss function during network training which consists of two terms, including a shape term and motion term.
\begin{equation}
L_{lip}=\sum_{t=1}^{t_n}\left \| b_t - \hat{b_t} \right \|_2 + \left \| (b_t-b_{t-1}) - (\hat{b_t}-\hat{b_{t-1}}) \right \|_2
\end{equation}
where $b_t$ is the predicted 3D facial parameters at frame $t$, while $\hat{b_t}$ is the corresponding ground truth. Besides, we further adopt articulation correction by the given phoneme label, for example, the mouth should be closed during the pronunciation of `b/p/m'. With the full exploration of both speech and phoneme information, we can generate accurate and synchronous lip movements.

\paragraph{Expression generation.}
The facial expression in the upper face mainly lies in the movements of eyes and eyebrows, which are related to speech rhythm and intention of the speaker with longer-time dependencies. 
Therefore, we divide the facial expression into two parts. First, we generate rhythmic expression movement by learning-based framework. Specifically, we employ an audio encoder to model the current speech signals as well as a motion encoder to model the history expressions, then a transformer decoder is adopted to predict the final expression movements according to audio and history motion information. Since synthesizing expression is a one-to-many mapping, we use SSIM loss to explore the structural similarity between the predicted expression and ground truth, which is defined as follows.
\begin{equation}
L_{exp}=1-\frac{(2\mu\hat{\mu}+\delta_1)(2cov+\delta_2)}{(\mu^2+\hat{\mu}^2+\delta_1)(\sigma^2+\hat{\sigma}^2+\delta_2)}
\end{equation}
where $\mu$ and $\sigma$ are the mean and standard deviation of generated 3D facial parameter sequence, while $cov$ is the covariance. Thus, we can generate rhythmic expression movement given the speech signals.

%Second, we apply intention-driven facial expression generation where the semantic tags are extracted from textual scripts according to the keywords, which indicate a specific intention such as emphasis. Besides, we  recorded more than 50 intention-based expressions as our emotion library, so that we can merge the generated facial movements with the pre-defined expressions to get the final expressive and diverse facial animation.
Second, we generate intention-driven facial expression based on the semantic tags which are extracted from textual scripts via sentiment analysis. The semantic tags include happiness, sadness, emphasis, fear and etc. Actors are asked to performers more than 50 intention-based expressions according to the semantic tags as our intention-driven expression database, so that we can generation facial expression by fusing the generated rhythmic expression with the proper intention-driven expression triggered by the semantic tags, and further integrate the lip movements to form the final expressive and diverse facial animation.

\subsection{Body Animation}
We adopt a graph-based approach to achieve fully automated body motion synthesis for the given speech and text. The proposed method is divided into two stages. First, we build a graph based on an existing motion database. Then the motion segments are retrieved from the graph according to the features of the given text/speech, and then concatenated into a long sequence.

\paragraph{Motion graph construction.}
To achieve high-quality motion synthesis, we invite professional actors to collect motion data through a Vicon Mo-Cap device and carefully fix corruptions in data. The database mainly contains two types of motions: semantic motions and non-semantic motions. Semantic motions include 24 kinds of actions(such as numbers, orientations, and special semantics) that have strong semantic correlations with the semantics of text/speech. Non-semantic motions are many declarative actions(upper body movements of standing, body center shifting movements, foot stepping movements, etc.) that can be used to match phonetic rhythms, and there is an one-to-many relationship between text/speech and 
animation.

A motion graph is a directed graph where a node denotes a motion segment and an edge denotes the cost of transition between two nodes.  The construction of  motion graph  includes: obtain graph nodes (dividing each long sequence in database to obtain many small motion segments), and build graph edges (constructing the connection relationship between motion segments). To obtain graph nodes, we extracted the motion strength of the long sequence, and take the frame indices of the local minima of the motion strength as dividing points. Then we build the edges between the two adjacent nodes by calculating the transition cost based on the distances between salient joint positions and movement speeds. A graph edge can be created if the transition cost between adjacent nodes is below a threshold $\sigma $. In particular, our database contains semantic motions, which must be complete motions to express semantics, so semantic motions in the motion graph need to be obtained manually.

%The motion speed near the dividing point is relatively slow, which ensures the motion stability of each graph node(motion clip) at the start and end frame, and helps to build the edges of the motion graph. The motion strength can be represented by the sum of the velocity of salient joints(head, left/right foot, knee, hand, elbow)~\cite{lee2019dancing}.

\paragraph{Graph-based retrieval and optimization.}
Synthesizing long sequence for the input text/speech needs to satisfy various rules (special semantic text and phonetic rhythm). For the motion graph, each synthesized long sequence corresponds to a path in the constructed motion graph. Our goal is to find optimal paths that satisfies the rules. Given a section of text/speech $P$, we first need to analyze the input, and divide it into many phrases ($P _{i}, i=1,...,n$) according to text structure and find the special semantic text in the section. We retrieve all meaningful motion segments from motion graph using the semantic text and the similarity of rhythm between motion segment and speech phrase (motion and phonetic rhythm are obtained by motion strength~\cite{lee2019dancing} and librosa~\cite{mcfee2015librosa}, respectively). To assign a motion node in the motion graph to each text/speech phrase so that the cost is minimized:
\begin{equation}
C=\lambda  _{t}\sum_{i=1}^{n-1}C_{t}(i,i+1)+\sum_{i=1}^{n-1}C_{r}(i)
\end{equation}

\begin{align}
C_{p}(i)=\left\{\begin{matrix}
&\lambda  _{s}C_{s}(i), \; &if \; P_{i} \;\; is\; the\; sematic\; text;\\ 
&\lambda  _{r}C_{r}(i), \; &otherwise
\end{matrix}\right.
\end{align}
where $C_{t}(i,i+1)$ is the transition cost between adjacent nodes, $C_{p}(i)$ accounts for the loss of special semantic text $C_{s}(i)$ and phonetic rhythm $C_{r}(i)$. $\lambda  _{t}, \lambda  _{s}, \lambda  _{r}$ are weights.

\subsection{Result}
We combine face and body movements, then skin and render to get the final video stream, as shown in Figure~\ref{fig:framework}. In the demo video, we show two demos(Chinese and English text/speech cases) for face and body animation respectively. All the results demonstrate that our system can synthesize natural, diverse and co-speech animations. More importantly, our system can generate different movements for the same text/speech, which fully embodies the one-to-many relationship between text/speech and animation. In addition, our system is sufficiently controllable and comprehensive for users to produce desired results, and the generated animations are accepted by professional artists and ordinary users. Therefore, our system can be applied to the Taobao virtual live for real-time live broadcast.

\section{Conclusion And Future Work}
In this paper, we propose a production-ready full-body animation synthesis system that takes text/speech as input and outputs face and body animation. Our system includes a novel learning-based method to generate lip movement and face expression, and a graph-based retrieval method to synthesize body motion. The generated results demonstrate that our system can robustly and efficiently generate realistic and high-quality full-body animation by given text/speech. In the future work, we will focus on the diverse emotions, personalities and interactivity to further enhance the expressiveness of 3D avatar.

\bibliographystyle{named}
\bibliography{ijcai22}

\begin{thebibliography}{}

\bibitem[\protect\citeauthoryear{Alexanderson \bgroup \em et al.\egroup
  }{2020}]{alexanderson2020style}
Simon Alexanderson, Gustav~Eje Henter, Taras Kucherenko, and Jonas Beskow.
\newblock Style-controllable speech-driven gesture synthesis using normalising
  flows.
\newblock In {\em Computer Graphics Forum}, volume~39, pages 487--496. Wiley
  Online Library, 2020.

\bibitem[\protect\citeauthoryear{Chen \bgroup \em et al.\egroup
  }{2019}]{chen2019hierarchical}
Lele Chen, Ross~K Maddox, Zhiyao Duan, and Chenliang Xu.
\newblock Hierarchical cross-modal talking face generation with dynamic
  pixel-wise loss.
\newblock In {\em Proceedings of the IEEE/CVF Conference on Computer Vision and
  Pattern Recognition}, pages 7832--7841, 2019.

\bibitem[\protect\citeauthoryear{Chen \bgroup \em et al.\egroup
  }{2021}]{chen2021choreomaster}
Kang Chen, Zhipeng Tan, Jin Lei, Song-Hai Zhang, Yuan-Chen Guo, Weidong Zhang,
  and Shi-Min Hu.
\newblock Choreomaster: choreography-oriented music-driven dance synthesis.
\newblock {\em ACM Transactions on Graphics (TOG)}, 40(4):1--13, 2021.

\bibitem[\protect\citeauthoryear{Ferstl \bgroup \em et al.\egroup
  }{2021}]{ferstl2021expressgesture}
Ylva Ferstl, Michael Neff, and Rachel McDonnell.
\newblock Expressgesture: Expressive gesture generation from speech through
  database matching.
\newblock {\em Computer Animation and Virtual Worlds}, 32(3-4):e2016, 2021.

\bibitem[\protect\citeauthoryear{Ginosar \bgroup \em et al.\egroup
  }{2019}]{ginosar2019learning}
Shiry Ginosar, Amir Bar, Gefen Kohavi, Caroline Chan, Andrew Owens, and
  Jitendra Malik.
\newblock Learning individual styles of conversational gesture.
\newblock In {\em Proceedings of the IEEE/CVF Conference on Computer Vision and
  Pattern Recognition}, pages 3497--3506, 2019.

\bibitem[\protect\citeauthoryear{Hu \bgroup \em et al.\egroup
  }{2021a}]{hu2021text}
Li~Hu, Jinwei Qi, Bang Zhang, Pan Pan, and Yinghui Xu.
\newblock Text-driven 3d avatar animation with emotional and expressive
  behaviors.
\newblock In {\em Proceedings of the 29th ACM International Conference on
  Multimedia}, pages 2816--2818, 2021.

\bibitem[\protect\citeauthoryear{Hu \bgroup \em et al.\egroup
  }{2021b}]{hu2021virtual}
Li~Hu, Bang Zhang, Peng Zhang, Jinwei Qi, Jian Cao, Daiheng Gao, Haiming Zhao,
  Xiaoduan Feng, Qi~Wang, Lian Zhuo, et~al.
\newblock A virtual character generation and animation system for e-commerce
  live streaming.
\newblock In {\em Proceedings of the 29th ACM International Conference on
  Multimedia}, pages 1202--1211, 2021.

\bibitem[\protect\citeauthoryear{Karras \bgroup \em et al.\egroup
  }{2017}]{karras2017audio}
Tero Karras, Timo Aila, Samuli Laine, Antti Herva, and Jaakko Lehtinen.
\newblock Audio-driven facial animation by joint end-to-end learning of pose
  and emotion.
\newblock {\em ACM Transactions on Graphics (TOG)}, 36(4):1--12, 2017.

\bibitem[\protect\citeauthoryear{Kovar \bgroup \em et al.\egroup
  }{2008}]{kovar2008motion}
Lucas Kovar, Michael Gleicher, and Fr{\'e}d{\'e}ric Pighin.
\newblock Motion graphs.
\newblock In {\em ACM SIGGRAPH 2008 classes}, pages 1--10. 2008.

\bibitem[\protect\citeauthoryear{Lee \bgroup \em et al.\egroup
  }{2019}]{lee2019dancing}
Hsin-Ying Lee, Xiaodong Yang, Ming-Yu Liu, Ting-Chun Wang, Yu-Ding Lu,
  Ming-Hsuan Yang, and Jan Kautz.
\newblock Dancing to music.
\newblock {\em Advances in Neural Information Processing Systems}, 32, 2019.

\bibitem[\protect\citeauthoryear{McFee \bgroup \em et al.\egroup
  }{2015}]{mcfee2015librosa}
Brian McFee, Colin Raffel, Dawen Liang, Daniel~P Ellis, Matt McVicar, Eric
  Battenberg, and Oriol Nieto.
\newblock librosa: Audio and music signal analysis in python.
\newblock In {\em Proceedings of the 14th python in science conference},
  volume~8, pages 18--25. Citeseer, 2015.

\bibitem[\protect\citeauthoryear{Oh \bgroup \em et al.\egroup
  }{2019}]{oh2019speech2face}
Tae-Hyun Oh, Tali Dekel, Changil Kim, Inbar Mosseri, William~T Freeman, Michael
  Rubinstein, and Wojciech Matusik.
\newblock Speech2face: Learning the face behind a voice.
\newblock In {\em Proceedings of the IEEE/CVF Conference on Computer Vision and
  Pattern Recognition}, pages 7539--7548, 2019.

\bibitem[\protect\citeauthoryear{Shlizerman \bgroup \em et al.\egroup
  }{2018}]{shlizerman2018audio}
Eli Shlizerman, Lucio Dery, Hayden Schoen, and Ira Kemelmacher-Shlizerman.
\newblock Audio to body dynamics.
\newblock In {\em Proceedings of the IEEE conference on computer vision and
  pattern recognition}, pages 7574--7583, 2018.

\bibitem[\protect\citeauthoryear{Wiles \bgroup \em et al.\egroup
  }{2018}]{wiles2018x2face}
Olivia Wiles, A~Koepke, and Andrew Zisserman.
\newblock X2face: A network for controlling face generation using images,
  audio, and pose codes.
\newblock In {\em Proceedings of the European conference on computer vision
  (ECCV)}, pages 670--686, 2018.

\end{thebibliography}

\end{document}